\documentstyle[12pt]{article}
\begin{document}
\centerline{\bf BEYOND THE SIMPLEST INFLATIONARY COSMOLOGICAL MODELS}
\bigskip
\centerline{A.A. Starobinsky}
\centerline{Landau Institute for Theoretical Physics,}
\centerline{Russian Academy of Sciences, Moscow 117334, Russia}

\vspace{0.5cm}

\noindent
Though predictions of the simplest inflationary cosmological models
with cold dark matter, flat space and approximately flat initial
spectrum of adiabatic perturbations are remarkably close to
observational data, we have to go beyond them and to introduce new
physics not yet discovered in laboratories to account for all data.
Two extensions of these models which seem to be the most actual at present
time are discussed. The first one is the possibility that we are living
at the beginning of a new inflation-like era. Then classical
cosmological tests, like the luminosity distance or the angular size
of distant objects as functions of redshift, as well as the behaviour of
density perturbations in a dustlike matter component including baryons
as a function of redshift, can provide information sufficient for the
unambiguous determination of an effective potential of a corresponding
present inflaton scalar field. The second, unrelated extension is a
possibility of broken-scale-invariant cosmological models which have
localized steps or spikes in the primordial perturbation spectrum. These
features can be produced by fast phase transitions in physical fields
other than an inflaton field in the early Universe during inflation and
not far from the end of it. At present, it seems that the only scale in
the spectrum around which we might see something of this type is
$k=0.05~h$ Mpc$^{-1}$.

\vspace{0.5cm}

\centerline{\bf 1. Introduction}

\vspace{0.5cm}

It is clear now that viable cosmological models of the present Universe
cannot be constructed without cold non-baryonic dark matter (CDM) and
some model of the early Universe producing an approximately flat
(or, Harrison-Zeldovich-like, $n_S\approx 1$) initial spectrum of
scalar (adiabatic) perturbations. The two necessary ingredients
correspond to physics not yet discovered in laboratories. In particular,
since the simplest cosmic defect models without inflation have failed
to fit observational data on the large-scale structure and angular
anisotropies of the cosmic microwave background (CMB) temperature in
the Universe, the only remaining viable and well elaborated scenario
producing such a spectrum is the inflationary scenario with one effective
scalar field (inflaton). The existence of the inflaton field with
desired properties of its self-interaction potential is a hypothesis about
physics at very high energies, only few orders of magnitude below the
Planck scale (in addition to this hypothesis about fundamental physics,
we have to assume that our Universe did really
pass through an inflationary stage in the past). Two main candidates
for non-baryonic CDM - the lightest supersymmetric particle or the
axion - are still awaiting a direct discovery of any (or, even both)
of them.

However, it has been already understood for several years that this amount
of new physics is not sufficient for cosmology. To explain all existing
observational data, one has to go further and to introduce even more
new physics. Really, it is well established that the simplest
Friedmann-Robertson-Walker (FRW) cosmological model with CDM, the flat
3-space ($\Omega_0=1$) and $n_S=1$ does not fit observational data.
Looking at viable extensions of this model presented in the Table 2
of the author's plenary talk at the "Cosmion-94" conference \cite{clas},
it is clear that all of them require new physics. It enters either at very
low energies -- producing an effective cosmological constant (the
$\Lambda$CDM model), or at medium energies -- giving a non-zero rest mass
to one or several neutrino types (CHDM models), or, finally, at very high
energies -- resulting either in phase transitions during single
(one-scalar-field) inflation
(a second-order phase transition in the case of broken-scale-invariant
(BSI) CDM models, or a first-order transition in the case of open CDM
(OCDM) models), or in the appearance of a second inflaton scalar field
(double inflation) that also lead to BSI CDM, or OCDM depending on
parameters.

After three years passed from the "Cosmion-94" conference, all these
four classes of models are still remaining viable in the sense that neither
of them may be excluded completely. However, observational evidence
(especially, the most recent one) becomes more and more directed
towards the $\Lambda$CDM scale-invariant model. That is why I consider it as
the model No. 1 at present, and I shall focus my main attention on it in
Sec. 2. Still there exist some data, however inconclusive they might
be, indicating that the present matter perturbation power spectrum has some 
feature at $k\approx 0.05~h$ Mpc$^{-1}$. If so, then it can be verified that 
this feature is present in the initial perturbation spectrum, too. It means
that we are dealing with a BSI CDM model. This possibility
does not remove need in a cosmological constant, so actually a BSI
$\Lambda$CDM model is required for a better fit to data. This topic is
considered in Sec. 3.

\vspace{0.5cm}

\centerline{\bf 2. Cosmological models with a variable cosmological term}

\vspace{0.5cm}

It has been known for many years that the flat
FRW cosmological model with cold dark matter, a positive cosmological
constant $\Lambda >0$ ($\Omega_0+\Omega_{\Lambda}=1$) and an approximately
flat spectrum of primordial adiabatic perturbations fits observational
data better and has a
larger admissible region of the parameters $(H_0,\Omega_0)$ than any other
cosmological model with both inflationary and non-inflationary initial
conditions (see, e.g., \cite{kof}). Here $H_0$ is the Hubble constant,
$\Omega_0=8\pi G\rho_m/3H_0^2$ includes baryons and (mainly)
non-baryonic dark matter, $\Omega_{\Lambda}\equiv\Lambda/3H_0^2$ and
$c=\hbar = 1$. This conclusion was based on the following
arguments: a) relation between $H_0$ and the age of the Universe $t_0$,
b) the fact that observed mass/luminosity ratio never leads to values
more than $\Omega_0 \sim 0.4$ up to supercluster scales, c) comparison
of cosmic microwave background temperature anisotropies,
power spectra of density and velocity matter perturbations,
present abundance of galaxy clusters with predictions of cosmological
models with inflationary initial conditions; d) observed values
of $\rho_b/\rho_m$ in rich galaxy clusters confronted with the range
for the present baryon density $\rho_b$ admitted by the theory of
primordial (Big Bang) nucleosynthesis. I don't include gravitational
lensing tests (e.g., a number of lensed quasars) here, since
conclusions based on them are less definite at present.

During last year two new pieces of strong evidence for $\Omega_0<1$
have appeared. The first (historically) of them is based on the evolution
of abundance of rich galaxy clusters with redshift $z$ \cite{bah}.
Still, it should be noted that there have been already appeared
some doubts on validity of the conclusion that $\Omega_0=1$ is really
excluded \cite{blan}. Much better observational data expected in
near future will help to resolve this dilemma unambiguously. The second,
completely independent argument for $\Omega_0=(0.2 - 0.4)$ follows from
direct observations of supernovae (type Ia) explosions at high redshifts
up to $z\sim 1$ \cite{garn}. On the other hand, no direct evidence
for a negative spatial curvature of the Universe (i.e., for the OCDM
model) has been found. Of course,
the possibility to have {\em both} a positive cosmological
constant and spatial curvature of any sign is not yet excluded, but,
according to the "Okkam's razor" principle, it would be desirable
not to introduce one more basic novel feature of the Universe
(spatial curvature) without conclusive observational evidence. In any
case, in spite of many theoretical and experimental attempts to
exorcize it, a $\Lambda$-term is back again.

It is clear that the introduction of a cosmological constant requires
new and completely unknown physics in the region of ultra-low energies.
Solutions with a cosmological constant occur in such
fundamental theories as supergravity and M-theory. However,
this cosmological constant is always negative and very large.
As compared to such a basic "vacuum" state, a very small and positive
cosmological constant allowed in the present Universe may be thought
as corresponding to the energy density $\varepsilon_{\Lambda}$ of a highly
excited (though still very symmetric) "background" state. So, it need not
be very "fundamental". But then it is natural to omit the assumption
that it should be exactly constant. In this case the name "a cosmological
term" (or a $\Lambda$-term) is more relevant for it, so I shall use this
one below. The principal difference between two kinds of non-baryonic
dark matter - dustlike CDM and a $\Lambda$-term - is that the latter
one is not gravitationally clustered up to scales $\sim 30~h^{-1}$
or more (otherwise we would return to the problem why $\Omega_0$ observed
from gravitational clustering is not equal to unity). Here $h=H_0/100$
km~s$^{-1}$Mpc$^{-1}$.

On the other hand, there exists a well-known strong argument showing that
a $\Lambda$-term cannot change with time as fast as the matter density
$\rho_m$ and the Ricci tensor (i.e., $\propto t^{-2}$) during the
matter dominated stage (for redshifs $z < 4\cdot 10^4~h^2$). Really, if
$\varepsilon_{\Lambda} \propto \rho_m$, so that $\Omega_{\Lambda}=const$,
then matter density perturbations in the CDM+baryon component grow as
$\delta\equiv \left({\delta \rho\over \rho}\right)_m \propto
t^{\alpha} \propto (1+z)^{-3\alpha/2}, ~\alpha= {\sqrt{25-24
\Omega_{\Lambda}}-1 \over 6}$. As a consequence, the total growth of
perturbations $\Delta$ since the time of equality of matter and radiation
energy densities up to the present moment is less than in the absence
of the $\Lambda$-term. If $\Omega_{\Lambda}\ll 1$, then
$\Delta (\Omega_{\Lambda})=\Delta(0)(1-(6.4+2\ln h)\Omega_{\Lambda})$.
Since parameters of viable
cosmological models are so tightly constrained that $\Delta$ may not be
reduced by more than twice approximately, this type of a $\Lambda$-term
cannot account for more than $\sim 0.1$ of the critical energy density
(see \cite{fer} for detailed investigation confirming this conclusion).
This, unfortunately, prevents us from natural explanation of the
present $\Lambda$-term with $\Omega_{\Lambda} = (0.5 - 0.8)$
using "compensation" mechanisms \cite{dolg} or exponential
potentials with sufficiently large exponents \cite{wet}.

A natural and simple description of a variable $\Lambda$-term is just
that which was so successively used to construct the simplest versions
of the inflationary scenario, namely, a scalar field with some interaction
potential $V(\varphi)$ minimally coupled to the Einstein gravity. Such an
approach, though phenomenological, is nevertheless more consistent and
fundamental than a commonly used attempt to decribe a $\Lambda$-term by a
barotropic ideal fluid with some equation of state. The latter approach
cannot be made internally consistent in case of negative pressure.
In particular, it
generally leads to imaginary values of the sound velocity. On the
contrary, no such problems arise using the scalar field description
(this scalar field is called the $\Lambda$-field below).
Of course, its effective mass $|m_{\varphi}^2|=|d^2V/d\varphi^2|$
should be very small to avoid gravitational clustering of this field in
galaxies, clusters and superclusters. To make a $\Lambda$-term
slowly varying, we assume that $|m_{\varphi}|\sim H_0 \sim 10^{-33}$ eV,
or less (though this condition may be relaxed). Models with a
time-dependent $\Lambda$-term were introduced more
than ten years ago \cite{ozer}, and different potentials $V(\varphi)$
(all inspired by inflationary models) were considered: exponential
\cite{wet,pib,fer,viana}, inverse power-law \cite{rat},
power-law \cite{weiss}, cosine \cite{friem,viana}.

However, it is clear that since we know essentially nothing about physics
at such energies, there exists no preferred theoretical candidate
for $V(\varphi)$. In this case, it is more natural to go from observations
to theory, and to determine an effective phenomenological potential
$V(\varphi)$ from observational data. The two new tests mentioned above
are just the most suitable for this aim. Really, using the cluster
abundance $n(z)$ determined from observations and assuming the Gaussian
statistics of initial perturbations (the latter follows from the paradigm
of one-field inflation, and it is in agreement with other observational
data), it is possible to determine a {\em linear} density perturbation
in the CDM+baryon dustlike component $\delta (z)$ for a fixed comoving scale
$R\sim 8(1+z)^{-1}h^{-1}$ Mpc up to $z\sim 1$, either using the
Press-Schechter approximation, or by direct numerical simulations
of nonlinear gravitational instability in the expanding Universe.
$\delta (z)$ can be also determined from observation of gravitational
clustering (in particular, of the galaxy-galaxy correlation function) as a
function of $z$. On the other hand, observations of SNe at different $z$
yield the luminosity distance $D_L(z)$ through the standard astronomical
expression $m=M+5\log D_L +25$, where $m$ is the observed magnitude,
$M$ is the absolute magnitude and $D_L$ is measured in Mpc. One more
related cosmological test is the angular size $\theta (z)$ of an
extended object with a proper size $d$. This test is less accurate at
present due to absence of objects with their size being fixed to a 
desired accuracy, still it is important.

Let us show now how to determine $V(\varphi)$ from either $\delta(z)$,
or $D_L(z)$, or $\theta(z)$. Also, we investigate what
additional information is necessary for an unambiguous solution of this
problem in all these cases.
The derivation of $V(\varphi)$ consists of two steps. First, the Hubble
parameter $H\equiv {\dot a\over a}=H(z)$ is determined. Here $a(t)$ is
the FRW scale factor, $1+z\equiv a_0/a$, the dot means ${d\over dt}$ and the
index $0$ denotes the present value of a corresponding quantity
(in particular, $H(t_0)=H(z=0)=H_0$). In the case of SNe, the first step
is almost trivial since the textbook expression for $D_L$ reads:
\begin{equation}
D_L(z) = a_0(\eta_0 -\eta)(1+z),~~\eta= \int_0^t{dt\over a(t)}~.
\end{equation}
Therefore,
\begin{equation}
H(z)={da\over a^2d\eta}=-(a_0\eta')^{-1}=
\left[\left({D_L(z)\over 1+z}\right)'\right]^{-1}~.
\label{DLH}
\end{equation}
Here and below in Sec. 2, a prime denotes the derivative with respect to $z$.

The angular size is given by
\begin{equation}
\theta (z) = {d\over a(\eta)(\eta_0 - \eta)}= {d\,(1+z)\over a_0(\eta_0 -
\eta)}~.
\end{equation}
Thus,
\begin{equation}
H(z)= - (a_0\eta')^{-1}= \left[d\left({1+z\over \theta(z)}\right)'
\right]^{-1}~.
\label{ang}
\end{equation}
In physical units, the right-hand sides of Eqs.~(\ref{DLH},~\ref{ang})
should be multipled by $c$. Thus, both $D_L(z)$ and $\theta (z)$ define
$H(z)$ uniquely.

More calculations are required to find $H(z)$ from $\delta(z)$. The
system of background equations for the system under consideration is:
\begin{equation}
H^2={8\pi G\over 3}\left(\rho_m + {\dot\varphi^2\over 2}+ V\right)~, ~~~
\rho_m={3\Omega_0H_0^2a_0^3\over 8\pi Ga^3}~,\label{00}
\end{equation}
\begin{equation}
\ddot \varphi +3H\dot\varphi +{dV\over d\varphi}=0~, \label{phieq}
\end{equation}
\begin{equation}
\dot H=-4\pi G(\rho_m + \dot\varphi^2)~. \label{alphaeq}
\end{equation}
Eq.~(\ref{alphaeq}) is actually the consequence of the other two equations.

We consider a perturbed FRW background which metric, in the
longitudinal gauge (LG), has the form:
\begin{equation}
ds^2=(1+2\Phi)dt^2-a^2(t)(1+2\Psi)\delta_{lm}dx^ldx^m, ~~l,m=1,2,3~.
\end {equation}
The system of equations for scalar perturbations reads (the spatial
dependence $\exp(ik_lx^l)$, $k_lk^l\equiv k^2$ is assumed):
\begin{equation}
\Phi=\Psi=\dot v~,~~~\dot\delta = -{k^2\over a^2}v +
3(\ddot v+H\dot v+\dot Hv)~,
\label{delta}
\end{equation}
\begin{equation}
\dot \Phi + H\Phi= 4\pi G(\rho_mv+\dot\varphi \delta \varphi)~,
\end{equation}
\begin{equation}
\left(-{k^2\over a^2}+4\pi G\dot\varphi^2\right)\Phi=4\pi G
(\rho_m\delta +\dot\varphi\dot{\delta\varphi}+3H\dot\varphi\delta
\varphi +{dV\over d\varphi}\delta\varphi)~,
\label{Phi}
\end{equation}
\begin{equation}
\ddot{\delta\varphi}+3H\dot{\delta\varphi}+\left({k^2\over a^2}+
{d^2V\over d\varphi^2}\right)\delta\varphi = 4\dot\varphi\dot\Phi -
2{dV\over d\varphi}\Phi~.  \label{dphieq}
\end{equation}
Eq.~(\ref{dphieq}) is the consequence of other ones. Here $v$ and
$\delta\varphi$ are, correspondingly, a velocity potential of a
dustlike matter peculiar velocity and a $\Lambda$-field perturbation
in LG, and $\delta$ is a {\em comoving} fractional matter density
perturbation (in this case, it coincides with
$\left({\delta \rho\over \rho}\right)_m$ in the synchronous gauge).
In fact, all these perturbed quantities are gauge-invariant.

Now let us take a comoving wavelength $\lambda = k/a(t)$ which is
much smaller than the Hubble radius $H^{-1}(t)$ up to redshifts $z\sim 5$.
This corresponds to $\lambda \ll 2000~h^{-1}$ Mpc at present. Then,
from Eq.~(\ref{dphieq}),
\begin{equation}
\delta\varphi \approx {a^2\over k^2} (4\dot\varphi\dot\Phi -
2{dV\over d\varphi}\Phi),~~|\dot\varphi\dot{\delta\varphi}|
\sim |{dV\over d\varphi}\delta\varphi|\sim {a^2H^4\over Gk^2}|\Phi|
\ll \rho_m|\delta|~.
\label{ineq}
\end{equation}
It is at this place where we heavily used the condition $|m_{\phi}|
\stackrel{<}{\sim} H_0$. In particular, it gives us a possibility
to estimate $|\dot \Phi|$ as $H|\Phi|$ and $|\dot {\delta \varphi}|$
as $H|\delta\varphi|$. On the other hand, it is clear that this condition
may be somewhat relaxed without destroying the last inequality in
Eq. (\ref{ineq}).
As a result, the $\Lambda$-field is practically unclustered at the scale
involved. Now the last of Eqs.~(\ref{delta}) and Eq.~(\ref{Phi})
may be simplified to:
\begin{equation}
\dot \delta = -{k^2\over a^2}v, ~~-{k^2\over a^2}\Phi=4\pi G\rho_m\delta~.
\end{equation}
Combining this with the first of Eqs.~(\ref{delta}), we return to
a well-known equation for $\delta$ in the absence of the $\Lambda$-field:
\begin{equation}
\ddot \delta +2H\dot \delta - 4\pi G\rho_m \delta =0~.
\label{del1}
\end{equation}

It is not possible to solve this equation analytically for an
arbitrary $V(\varphi)$. Remarkably, the inverse dynamical problem,
i.e. the determination of $H(a)$ given $\delta(a)$, is solvable. After
changing the argument in Eq.~(\ref{del1}) from $t$ to $a$ (${d \over dt}=
aH{d\over da}$), we get a first order linear differential equation for
$H^2(a)$:
\begin{equation}
a^2{d\delta\over da}{dH^2\over da}+2\left(a^2{d^2\delta\over da^2}+
3a{d\delta\over da}\right)H^2={3\Omega_0H_0^2a_0^3\delta\over a^3}~.
\end{equation}
The solution is:
\begin{equation}
H^2={3\Omega_0H_0^2a_0^3\over a^6}\left({d\delta\over da}\right)^{-2}
\int_0^a a\delta {d\delta\over da}\, da = 3\Omega_0H_0^2{(1+z)^2\over
\delta'^2}\int_z^{\infty}{\delta |\delta'|\over 1+z}\, dz~.
\label{delH}
\end{equation}
Putting $z=0$ in this expression for $H$, we arrive to the
expression of $\Omega_0$ through $\delta (z)$:
\begin{equation}
\Omega_0=\delta'^2(0)\left(3\int_0^{\infty}{\delta |\delta'|
\over 1+z}\, dz\right)^{-1}~.
\label{omega}
\end{equation}
Of course, observations of gravitational clustering can hardly
provide the function $\delta (z)$ for too large $z$ (say, for $z>5$).
However, $\delta (z)$ in the integrands in Eqs.~(\ref{delH},\ref{omega})
may be well approximated by its $\Omega_0=1$ behaviour (i.e.,
$\delta \propto (1+z)^{-1}$) already for $z>(2-3)$. If massive
neutrinos are present, one should use here the expression with $\alpha$
written above and with $\Omega_{\Lambda}$ substituted by $\Omega_{\nu}/
\Omega_0$ (it is assumed that $\rho_m$ includes massive neutrinos, too).

Using Eq.~(\ref{omega}), Eq.~(\ref{delH}) can be represented
in a more convenient form:
\begin{equation}
{H^2(z)\over H^2(0)}={(1+z)^2\delta'^2(0)\over \delta'^2(z)}
- 3\Omega_0{(1+z)^2\over\delta'^2(z)}\int_0^z{\delta |\delta'|
\over 1+z}\, dz~.
\end{equation}
Thus, $\delta(z)$ uniquely defines the ratio $H(z)/H_0$. Of course,
appearance of derivatives of $\delta(z)$ in these formulas shows that
sufficiently clean data are necessary, but one may expect that such
data will soon appear. Let us remind also that, for $\Lambda\equiv
const$ ($V(\varphi)\equiv const$), we have
\begin{equation}
H^2(z)=H_0^2(1-\Omega_0+\Omega_0(1+z)^3),~~q_0\equiv -1 +
\left({d\ln H\over d\ln (1+z)}\right)_{z=0}={3\over 2}\,\Omega_0-1~,
\end{equation}
where $q_0$ is the acceleration parameter.

The second step - the derivation of $V(\varphi)$ from $H(a)$ - is very
simple. One has to rewrite Eqs.~(\ref{00},~\ref{alphaeq}) in terms of $a$
and take their linear combinations:
\begin{eqnarray}
8\pi GV(\varphi)=aH{dH\over da}+3H^2-{3\over 2}\Omega_0H_0^2
\left({a_0\over a}\right)^3~, \nonumber \\
4\pi Ga^2H^2\left({d\varphi\over da}\right)^2=-aH{dH\over da} -
{3\over 2}\Omega_0H_0^2\left({a_0\over a}
\right)^3~,
\label{V}
\end{eqnarray}
and then exclude $a$ from these equations, since the second of
Eqs.~(\ref{V}) is integrated trivially.

To show explicitly how the second step proceeds for a given $H(z)$,
let us consider the case when a cosmological term mimics the
negative spatial curvature dynamically, while the real spatial geometry
of the Universe remains flat. That is, we assume that
\begin{equation}
H^2(z)=H_0^2\left(\Omega_0(1+z)^3+(1-\Omega_0)(1+z)^2\right)~.
\end {equation}
Integration of the second of Eqs.~(\ref{V}) gives
\begin{equation}
{a\over a_0}={\Omega_0\over 1-\Omega_0}\sinh^2\left(\sqrt{\pi G}(\varphi
-\varphi_0+\varphi_1)\right)~,~~ \varphi_1={1\over\sqrt{\pi G}}\ln \left(
{\sqrt{1-\Omega_0}+1 \over \sqrt{\Omega_0}}\right)~.
\end{equation}
Substituting this into the first of Eqs.~(\ref{V}), we find the expression
for the effective potential:
\begin{equation}
V(\varphi) = {(1-\Omega_0)^3H_0^2\over 4\pi G\Omega_0^2}~
{1\over \sinh^4\left(\sqrt{\pi G}(\varphi -\varphi_0 +\varphi_1)\right)}~.
\end{equation}
At early times at the matter dominated stage, this potential diverges
$\propto (\varphi -\varphi_0 +\varphi_1)^{-4}$.

Therefore, the model of a $\Lambda$-term considered in this paper can
account for {\em any} observed forms of $D_L(z)$, $\theta(z)$ and
$\delta(z)$ which,
in turn, can be transformed into a corresponding effective potential
$V(\varphi)$ of the $\Lambda$-field. The only condition is that the
functions $H(z)$ obtained by these three independent ways should coincide
within observational errors. $D_L(z)$ and $\theta(z)$ uniquely determine
$V(\varphi)$, if
$\Omega_0$ is given additionally (the latter is required at the second
step, in Eqs.~(\ref{V})). $\delta(z)$ uniquely determines $V(\varphi)$
up to the multiplier $H_0^2$, the latter has to be given additionally
to fix an overall amplitude.

Observational tests which can falsify this
model do exist. In particular, a contribution to large-angle
${\Delta T\over T}$ CMB temperature anisotropy due to the integrated
(or, non-local) Sachs-Wolfe effect presents a possibility
to distinguish the model from more complicated models, e.g., with
non-minimal coupling of the $\Lambda$-field to gravity or to CDM.
However, the latter test is not an easy one, since this contribution
is rather small and partially masked by cosmic variance.

Still it may appear finally that the cosmological term is really
constant: $V(\varphi)=V_0=\varepsilon_{\Lambda}=const,~\varphi=
\varphi_0=const$. Then it represents a new fundamental constant of
nature. A question is often asked: how is it possible to obtain
such a small value of $\varepsilon_{\Lambda}$ in Planck units from
known physical constants? This
question is actually not physical, but an arithmetic one. So, without 
trying to construct an underlying physical model for the $\Lambda$-term,
let me propose a toy arithmetic expression for $\varepsilon_{\Lambda}$
(of course, one of infinite number of possible ones) to show that
there is no problem here from the mathematical point of view:
\begin{equation}
\varepsilon_{\Lambda} = {M_P^4\over (2\pi^2)^3}\exp\left(-{2\over \alpha}
\right) = e^{-283.02}M_P^4=10^{-122.91}M_P^4~,
\label{pred}
\end{equation}
where $M_P=\sqrt G$ is the Planck mass and $\alpha$ is the fine-structure
constant. In usual units, $\varepsilon_{\Lambda}=10^{-123}c^5/G^2\hbar$.
Eq.~(\ref{pred}) leads to the "prediction":
\begin{equation}
\Omega_{\Lambda}h^2\equiv (1-\Omega_0)h^2\approx 0.335~.
\end{equation}
In particular, for $\Omega_0=0.3$, it gives $H_0=69$ km~s$^{-1}$Mpc$^{-1}$
-- just in the expected region!

\vspace{0.5cm}

\centerline{\bf 3. A feature in the matter power spectrum}
\centerline{\bf and physical mechanisms to produce it}

\vspace{0.5cm}

The main importance of the inflationary scenario of the early Universe
for the theory of large scale structure in the present Universe is that
the former scenario predicts (in its simplest realizations) an approximately
flat, or scale-invariant, spectrum ($n_S(k)\equiv {d\ln P_0(k)/
d\ln k}\approx 1$) of initial adiabatic perturbations.
By the simplest realizations I mean, as usually, inflationary models with
one effective slow-rolling scalar (inflaton) field. Of course, the physical
nature of the inflaton may be completely different in these models, but it
does not matter for observations, in particular, for the large scale
structure. This prediction has been confirmed already, if by $n_S(k)$ we
understand the slope of an initial power spectrum of density
perturbations $P_0(k)$ smoothed over the range
$\Delta \ln k \sim 1$. To see this, it is even not neccessary to use
results for $n_S$ following from the COBE experiment (though they also tell
us the same), it is sufficient to compare the COBE normalization of
perturbations for scales of the order of the present cosmological horizon
$R_h$ with the $\sigma_8$ normalization that follows, e.g., from the present
cluster abundance. The difference in the amplitude of initial perturbations
at these two scales which are divided by approximately $3$ orders of
magnitude is only $2-2.5$ times for the standard CDM model and even less for
other models, e.g., the $\Lambda$CDM model. In addition, these numbers give
us an idea about the magnitude of expected deviations from the exact
scale invariance: $|n_S-1|\le 0.3$ (once more, we are speaking about a
smoothed $n_S$). Observational effects related to the part of $P_0(k)$
between these two points (CMB temperature fluctuations at
medium and small angles, galaxy-galaxy and cluster-cluster correlations,
peculiar velocities of galaxies) also do not require larger smooth deviations
from $n_S=1$.

The observational fact that the smoothed slope $n_S$ cannot be significantly
different from 1 does not exclude the possibility of {\em local} strong
deviations from the flat spectrum, i.e., steps and/or spikes in $P_0(k)$.
Of course, one should not expect such a behaviour to be typical,
we shall see below that if it happens at all, it occurs at some preferred
scales which themselves become new fundamental parameters
of a cosmological model. Do we have any observational evidence for an
existence of such preferred scales at the Universe? At present, only one
scale in the Fourier space, $k=k_0=0.05h$ Mpc$^{-1}$, remains a candidate for
this role, and it seems that the spectrum is smooth for larger $k$ (from
galaxy-galaxy correlation data) and for smaller $k$ (from CMB data). As for
this scale itself, there exists an evidence for a peculiar behaviour (in the
form of a sharp peak) in the Fourier power spectrum of rich Abell - ACO
clusters (with richness class $R\ge 0$ and redshifts $z\le 0.12$) around it
\cite{ein} (see also \cite{ein1,ein2}). This anomalous behaviour persists
if the distant border for the cluster sample is reduced to $z=0.07-0.08$
\cite{retz}. If we assume that the
cluster power spectrum is proportional to the power spectrum of whole
matter in the Universe (with some constant biasing factor), and calculate
corresponding {\em rms} multipole values $C_l$ of angular fluctuations
of the CMB temperature, they appear to be in a good agreement with existing
results of medium-angle experiments \cite{atrio}. Moreover, if $\Omega_m=1$,
the peak in the power spectrum inferred from the cluster data just explains
an excess in $C_l$ for $l=200-300$ observed in the Saskatoon experiment.
The agreement becomes even better in the case of the $\Lambda$CDM model
\cite{davs}.

On the other hand, there is no peak at $k=k_0$ in the power spectrum of both
APM clusters (which are generally less rich than Abell-ACO clusters) and
APM galaxies \cite{tad} (though some less prominent feature at this scale
may still exist in the latter spectrum \cite{gast}), and the maximum in these
spectra seems to be shifted to $k\sim 0.03h$ Mpc$^{-1}$. Leaving a solution
of this discrepancy to more complete future surveys, let us consider
theoretical predictions.

It is possible to produce local features in the initial spectrum even
remaining (at least, formally) inside the standard paradigm of one-field
inflation. The only thing which should be relaxed is the requirement of the
analyticity of an inflaton effective potential $V(\phi)$ at all points
(the $\phi$ inflaton field in the early Universe should not be confused
with the $\varphi$ $\Lambda$-field used to describe a variable
cosmological term in the previous section).

So, let me admit that $V(\phi)$ has some kind of discontinuity at a point
$\phi=\phi_0$. Of course, really this discontinuity is smoothed in a
very small vicinity of $\phi_0$. Three cases are the most interesting.

1. $[V]=[V']=0,~[V'']\not= 0$ at $\phi=\phi_0$.

\noindent
Here $[~]$ means the jump in the quantity considered, namely,
$[A]\equiv A(\phi_0 +0)-A(\phi_0-0)$, and the prime denotes the
derivative
with respect to $\phi$. If we assume that the slow-roll
conditions $V'^{2}\ll 48\pi GV^2,~|V''|\ll 24\pi GV$ are satisfied near the
point $\phi=\phi_0$, then
in the zero-order approximation the standard result for a perturbation
spectrum is valid:
\begin{equation}
P_0(k)={k^4R_h^4(t)\over 400}h^2(k),~~k^3h^2(k)= 18\left({H^6\over V'^2}
\right)_k, ~~H\equiv {\dot a\over a}\approx \sqrt{{8\pi GV\over 3}},
\label{stand}
\end{equation}
where the index $k$ means that the quantity is evaluated at the moment of the
first Hubble radius crossing ($k=aH$) at the inflationary stage. The result
for $P_0(k)$, in contrast to the metric perturbation $h^2(k)$ defined in the
ultra-synchronous gauge ($h$ is equal to $1/3$ of the trace of a spatial
metric perturbation in this gauge), refers to the matter-dominated stage
where $R_h(t)=2/H=3t$. Note also that there is no necessity in adding the
multiplier ${\cal O}(1)$ here.

So, in this case $P_0(k)$ is continuous at $k=k_0$ but its slope $n_S(k)$
has a step-like behaviour there (similar to the case considered in
\cite{ein1,retz}). However, due to small corrections to Eq. (\ref{stand})
which are beyond the slow-roll approximation, it appears that $n_S$ cannot
be obtained simply by differentiating (\ref{stand}), and I expect that the
sharp behaviour in $n_S$ will be smoothed near $k_0$. This question is
still under consideration.

2. $[V]=0,~[V']\not= 0$ at $\phi=\phi_0$.

\noindent
Now the second of the slow-roll conditions is violated, while we can choose
parameters of the jump in such a way that the first condition is still valid.
Naive application of Eq.(\ref{stand}) would give a step in $P_0(k)$. However,
the slow-roll approximation is clearly not applicable. The exact solution for
a local part of the spectrum near the point $k_0$ was obtained in
\cite{star92}. It reads:
$$k^3h^2(k)={18H_0^6\over V_-'^2}D^2(y),~H_0=\sqrt{{8\pi GV(\phi_0)
\over 3}},~V'_{\pm}=V'(\phi_0\pm 0)>0,~y={k\over k_0}, $$
\begin{equation}
D^2(y)= 1-3\left({V_-'\over V_+'}-1\right){1\over y}\left(\left(1-
{1\over y^2}\right)\sin 2y+{2\over y}\cos 2y\right)
\label{step}
\end{equation}
$$+{9\over 2}\left({V_-'\over V_+'}-1\right)^2{1\over y^2}\left(1+
{1\over y^2}\right)\left(1+{1\over y^2}+\left(1-{1\over y^2}\right)\cos 2y-
{2\over y}\sin 2y\right). $$
The function $D^2(y)$ has a step-like behaviour with superimposed
oscillations.
Since $D(0)=V_-'/V_+',~D(\infty)=1$, the spectrum approaches the flat
spectrum
if $|\ln(k/k_0)|\gg 1$. As compared with the flat spectrum, the spectrum
(\ref{step}) has more power at large scales (small $k$) if $V_-'>V_+'$,
and more power on small scales in the opposite case. The shape of this
function is universal (in the sence that it does not depend on a way of
smoothing the jump in $V'$, if it is made in a sufficiently small vicinity of
$\phi_0$), it depends on the ratio $V_-'/V_+'$ only.

In the absence of a cosmological term, we have to choose the case
$V_-'>V_+'$ to fit observational data, and then it is not possible
to have a significant bump in the spectrum. However, if the cosmological
term is positive, a possibility of an inverted step ($V_-'<V_+'$) arises
\cite{davs}. In this case, the spectrum (\ref{step}) can match the spectrum
found in \cite{ein} rather well.

3. $[V] \not= 0$ at $\phi=\phi_0$.

\noindent
In this case, there is no universal spectrum, and the answer depends on a
concrete way of smoothing $V(\phi)$ at $\phi =\phi_0$. Some
general results
are presented in \cite{star92}, typically $P_0(k)$ acquires a large bump,
however, a well may appear, too. Fortunately, there is no need in further
consideration of this more complicated case, since observational data do
not require such a strong non-analyticity. Looking at the results presented
in \cite{ein,retz} it is clear that they lie somewhere between the first two
cases.

Therefore, a general lesson from these considerations is that pecularities in
$V(\phi)$ can produce local features in $P_0(k)$ where the slope $n_S$ is
significantly different from unity. However, these features cannot be too
sharp, in particular, both $P_0(k)$ and $n_S(k)$ are expected to be
continuous functions of $k$.

So far, the treatment of peculiar points was purely mathematical. However,
if we are seeking for a physical explanation of such behaviour of
$V(\phi)$,
we have to go beyond the paradigm of one-field inflation to a more
complicated case of two-field inflation. Inflation with two scalar field is
a very rich physical model which includes double inflation, hybrid inflation,
open inflation, etc. as specific cases. In our case it is sufficient to
assume that the second scalar field $\chi$, in contrast to the inflaton
field $\phi$, is always in the regime $|m_{\chi}^2|\gg H^2$. So, it is not
dynamically important during the whole inflation. However, it is coupled to
$\phi$ (e.g., through the term $g^2\phi^2\chi^2$) and, as a result
of change in $\phi$ during inflation, the field $\chi$ experiences a fast
phase transition approximately $60$ e-folds before the end of inflation.
If parameters of an interaction potential $V(\phi,\chi)$ are such that the
phase transition may be considered as an equilibrium one, then its net effect
on inflation appears in the change of the equilibrium effective potential
$V_{eff}(\phi)\equiv V(\phi,\chi_{eq}(\phi))$ only. If the
transition is a second-order
one, with no jump in $\chi_{eq}$, the first case considered above takes
place. If the transition is a first-order one, with a non-zero jump in
$\chi_{eq}$, we arrive to the second case.

This research was partially supported by the Russian Foundation for
Basic Research, Grant 96-02-17591, and by the Russian Research Project
"Cosmomicrophysics".

\vfill
\end{document}